\documentclass[prl,twocolumn,superscriptaddress,floatfix,nopacs]{revtex4-1}
\usepackage{graphicx,amsfonts,amssymb,amsmath,hyperref,enumerate,upgreek}

\newif\ifhyper
\hypertrue
\ifhyper
\hypersetup{
   citecolor = {green},
   colorlinks = {true}, 
   urlcolor = {blue} 
}
\fi

\newcommand{\beq}{\begin{equation}}
\newcommand{\eeq}{\end{equation}}
\newcommand{\beqa}{\begin{eqnarray}}
\newcommand{\eeqa}{\end{eqnarray}}
\newcommand{\ket} [1] {\vert #1 \rangle}

\newcommand{\tr}[2][]{\text{Tr}_{#1}\left\{#2\right\}}

\def\ket#1{\vert#1\rangle}
\def\ketbra#1{\vert#1\rangle\langle#1\vert}

\def\Longarrow{\protect\@lra}
\def\@lra{\relbar\joinrel\relbar\joinrel\relbar\joinrel%
          \relbar\joinrel\rightarrow}

\begin{document}

\title{Initialization of quantum simulators by sympathetic cooling}

\author{Meghana Raghunandan}
\email{meghana.raghunandan@itp.uni-hannover.de}
\affiliation{Institut f\"ur Theoretische Physik, Leibniz Universit\"at Hannover, Appelstra{\ss}e 2, 30167 Hannover, Germany}
\author{Fabian Wolf}
\affiliation{QUEST Institut, Physikalisch-Technische Bundesanstalt, Bundesallee 100, 38116 Braunschweig, Germany}
\author{Christian Ospelkaus}
\affiliation{QUEST Institut, Physikalisch-Technische Bundesanstalt, Bundesallee 100, 38116 Braunschweig, Germany}
\affiliation{Institut f\"ur Quantenoptik, Leibniz Universit\"at Hannover, Welfengarten 1, 30167 Hannover, Germany}
\author{Piet O. Schmidt}
\affiliation{QUEST Institut, Physikalisch-Technische Bundesanstalt, Bundesallee 100, 38116 Braunschweig, Germany}
\affiliation{Institut f\"ur Quantenoptik, Leibniz Universit\"at Hannover, Welfengarten 1, 30167 Hannover, Germany}
\author{Hendrik Weimer}
\affiliation{Institut f\"ur Theoretische Physik, Leibniz Universit\"at Hannover, Appelstra{\ss}e 2, 30167 Hannover, Germany}

\begin{abstract}

  Simulating computationally intractable many-body problems on a
  quantum simulator holds great potential to deliver insights
  into physical, chemical, and biological systems.  While the
  implementation of Hamiltonian dynamics within a quantum simulator
  has already been demonstrated in many experiments, the problem of
  initialization of quantum simulators to a suitable quantum state has
  hitherto remained mostly unsolved. Here, we show that already a
  single dissipatively driven auxiliary particle can efficiently
  prepare the quantum simulator in a low-energy state of largely
  arbitrary Hamiltonians. We demonstrate the scalability of our
  approach and show that it is robust against unwanted sources of
  decoherence. While our initialization protocol is largely
  independent of the physical realization of the simulation device, we
  provide an implementation example for a trapped ion quantum
  simulator.

\end{abstract}

\maketitle

\section{Introduction}

Quantum simulation is an emergent technology that can potentially
solve important open problems related to high-temperature
superconductivity, interacting quantum field theories, or many-body
localization \cite{Georgescu2014}. While a series of experiments
demonstrated the successful implementation of Hamiltonian dynamics
within a quantum simulator
\cite{Greiner2002,Schneider2008,Jordens2008,Friedenauer2008,Lanyon2011,Aidelsburger2013,Miyake2013,Alvarez2015,Mazurenko2017,Bernien2017,Zhang2017,Guardado-Sanchez2018,Lienhard2018},
these works had the simulator initialized in an easily accessible
state such as a product state. Consequently, adiabatic evolution from
an initial Hamiltonian whose ground state can be prepared to the
final Hamiltonian of interest has been used. However, this approach
becomes challenging across quantum phase transitions, especially if
the transition is of first order.

Our strategy to overcome this problem builds on the recent advances in
using dissipative quantum systems to engineer interesting many-body
states as the attractor states of such an open quantum many-body
system
\cite{Diehl2008,Verstraete2009,Weimer2010,Barreiro2011,Carr2013a,Rao2013,Lin2013,Shankar2013,Cormick2013,Morigi2015,Neto2017}. In
the past, these dissipative state engineering schemes have been
limited to ground states of stabilizer or frustration-free
Hamiltonians \cite{Verstraete2009,Weimer2010,Weimer2011,Roghani2018},
whose ground state can be found by performing local optimizations
alone. Unfortunately, almost all many-body Hamiltonians of interest
lie outside this class, requiring generalization of the dissipative state
preparation procedure.

\begin{figure}[t]
\includegraphics[width=\linewidth]{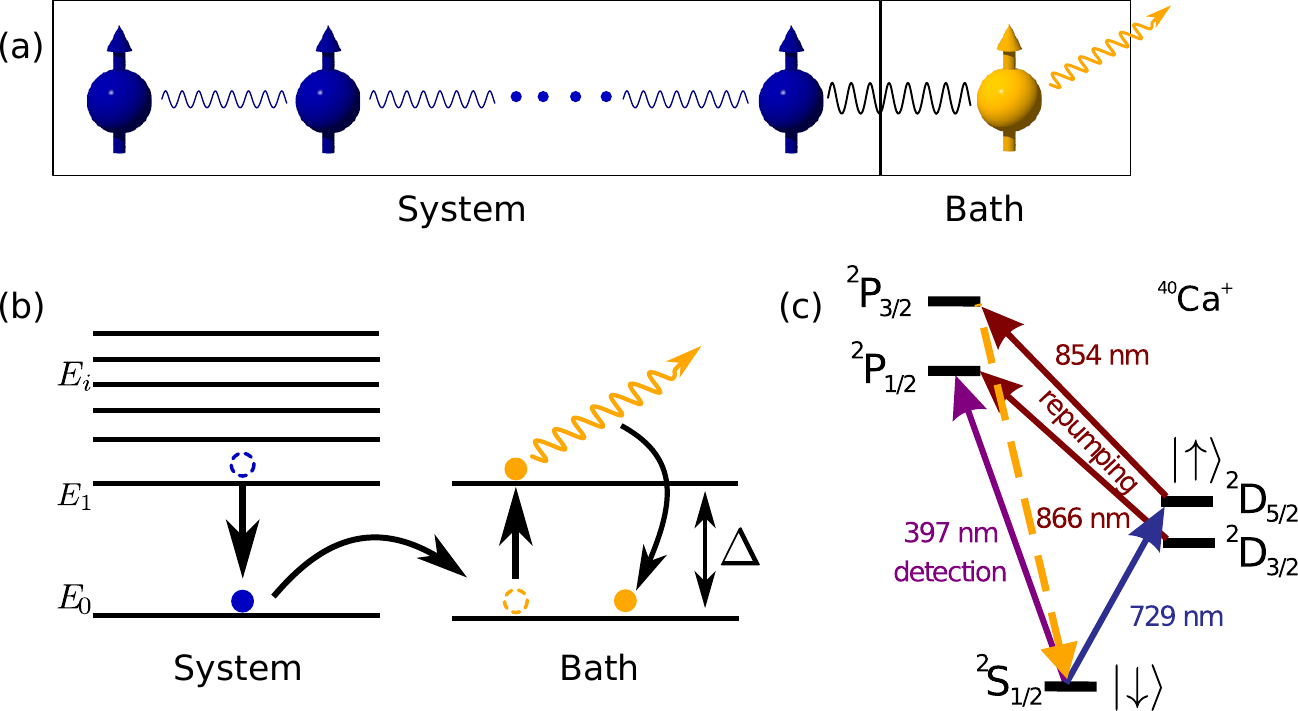}

\caption{Sympathetic cooling of a quantum simulator. (a) A system of $N$ spins performing the quantum simulation is interacting with an additional bath spin that is dissipatively driven. (b) Sketch of the energy level structure showing resonant energy transport between the system and the bath, after which the bath spin is dissipatively pumped into its ground state. (c) Level scheme for the implementation with trapped $^{40}$Ca$^+$ ions.}

\label{fig:setup}

\end{figure}

%
%
%

Here, we present a previously unexplored paradigm for the dissipative
initialization of a quantum simulator. We consider a coupling of the
many-body system performing the quantum simulation to an auxiliary
particle that is dissipatively driven. Crucially, the energy splitting
within the auxiliary particle is chosen such that it becomes resonant
with the many-body excitation gap of the system of interest, i.e., the
difference of the ground-state energy and the energy of the first
excited state. Under such a resonance condition, the energy of the
quantum simulator is efficiently transferred to the auxiliary particle
such that the former is cooled sympathetically
\cite{Cormick2013,Neto2017}. Although this setup is only resonant at a
single energy, the density of states increases exponentially with
energy, resulting in the lowest-lying excitations being the bottleneck
for fast ground-state preparation, \textcolor{black}{see the Supplementary Materials for details}. While the value of the many-body
excitation gap is usually unknown before performing the simulation, we
demonstrate that the gap can actually be determined from the quantum
simulation data in a spectroscopic measurement. Hence, the dissipative
initialization process provides important information about the
many-body system of interest at the same time. Notably, we show
that the cooling by a single auxiliary particle is efficient, and it
is especially robust against unwanted noise processes occurring in the
quantum simulator.

\begin{figure*}[bt]
\begin{tabular}{p{0.5cm}p{8.5cm}p{0.5cm}p{8.5cm}}
(a) & \vspace{-0.75cm} \includegraphics[width=8.5cm]{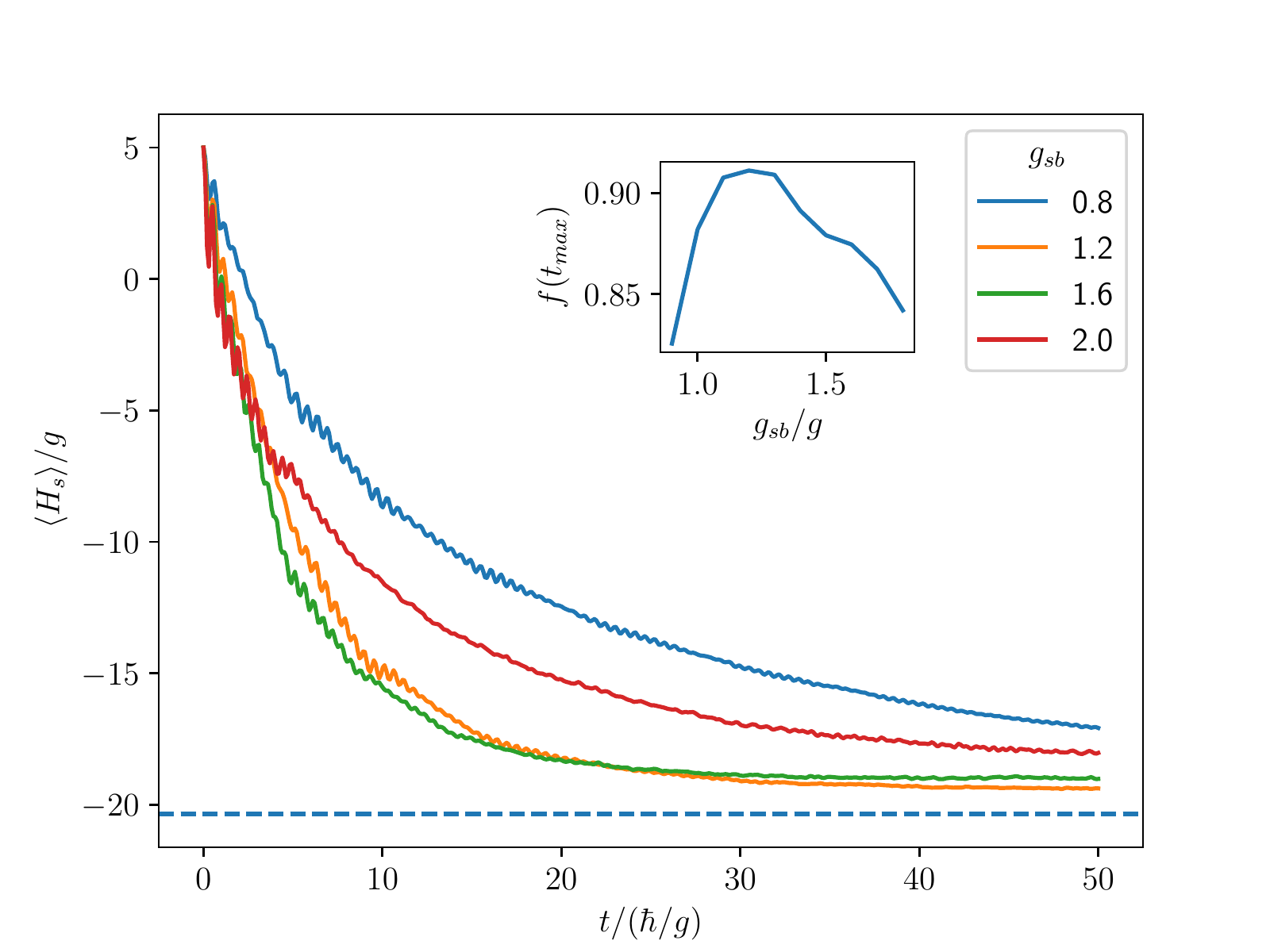} & (b) & \vspace{-.75cm}   \includegraphics[width=8.5cm]{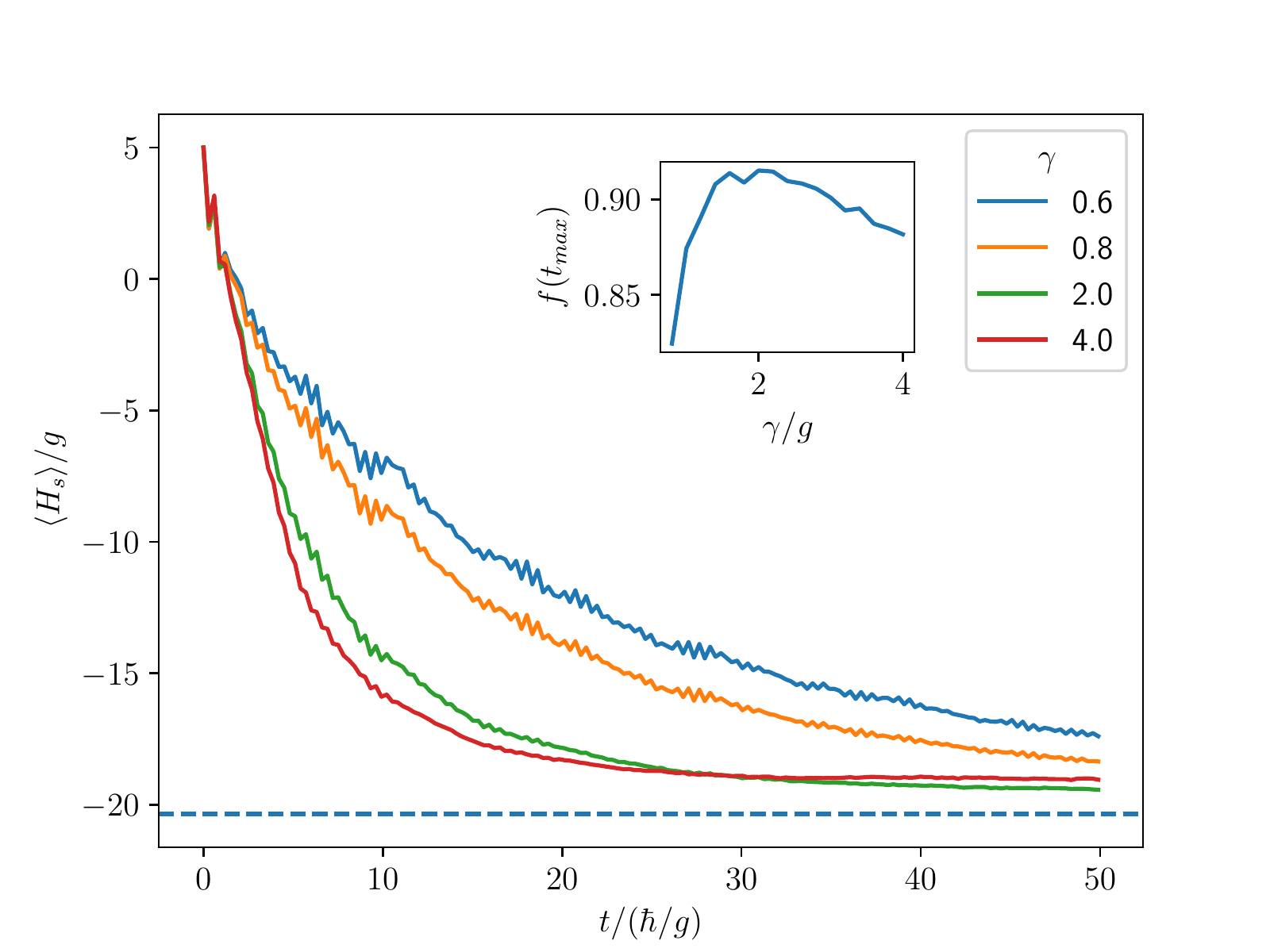}\\
\end{tabular}
\caption{Sympathetic cooling of the transverse field Ising model in
  the ferromagnetic phase ($J/g = 5$, $N=5$, \textcolor{black}{$f_{x,y,z}=\{ 1, 1.1, 0.9 \}$}). The speed of the cooling
  dynamics as well as the final energy of the system depend on the
  system-bath coupling $g_{sb}$ for $\gamma/g=1.9$ (a) and the dissipation rate $\gamma$ for $g_{sb}/g = 1.15$
  (b). The ground state energy is indicated by the dashed line. The
  insets show that the ground state can be prepared with greater than
  90\% fidelity.}
\label{fig:hsvspar}
\end{figure*}

To be explicit, we consider different paradigmatic one-dimensional
(1D) spin $1/2$ many-body systems coupled to a single dissipatively
driven auxiliary bath spin (see Fig.~\ref{fig:setup}). This setup can be
readily generalized to bosonic or fermionic many-body systems with a
larger local Hilbert space, to settings incorporating several bath
particles, and to higher spatial dimensions. In the following,
we assume a 1D chain of $N$ spins governed by the Hamiltonian
$H_{sys}$. One boundary spin of the system is coupled to the auxiliary
bath spin via an interaction Hamiltonian of the form $H_{int} = g_{sb}
\sum\limits_{x,y,z} f_i \sigma_{i}^{(N)}\sigma_{i}^{(b)}$, where
$g_{sb}$ is the strength of the system-bath interaction and the
$\sigma_i$ refer to Pauli matrices. The \textcolor{black}{exact values} of the dimensionless parameters $f_i$ \textcolor{black}{are} not particularly important. \textcolor{black}{In the models studied here, we find that it is either favorable to choose them roughly equal or have one dominant contribution. In addition, to avoid any symmetries in the interaction preventing the cooling of certain degrees of freedom, it is beneficial to assign slightly different values to them.}

The Hamiltonian of the bath spin $H_{bath}$ is given by $H_b = (\Delta/2)
\sigma_z^{(b)}$. The dissipation channel acting on the bath spins
performs dissipative spin flips from the up spin state to the down
spin state occurring with a rate $\gamma$. Then, the total dynamics is described by a quantum master equation in Lindblad form
\begin{equation}
\label{qme}
  \frac{d}{dt}\rho = -\frac{i}{\hbar}[H,\rho] + \gamma \left(\sigma_-^{(b)}\rho \sigma_+^{(b)} - \frac{1}{2}\left\{\sigma_+^{(b)}\sigma_-^{(b)}, \rho\right\}\right),
\end{equation}
where $H=H_{sys}+ H_{bath}+ H_{int} $ is the total Hamiltonian of the $N+1$
spin system \cite{Breuer2002}.

We would like to stress that such a setup imposes only modest
requirements for an experimental implementation,
\textcolor{black}{which works equally well for both analog and digital
  quantum simulators.} In particular, we note that our setup does not
require control over individual particles of the quantum simulator. In
our case, it is sufficient to merely be able to control the bath
particle independently of the rest of the system. In addition, the
dissipative dynamics can be induced by measuring the spin state of the
bath spin followed by a spin flip conditional on measuring the spin in
the up state.

In the methods
section, we give a detailed implementation guide for a trapped ion
quantum simulator.

\section{Results}

\subsection{Ising chain in a transverse field}

As the first paradigmatic model, we consider the Ising model in a
transverse field, given by the Hamiltonian
\begin{equation}
H_{sys} = g \sum\limits_{i=1}^{N} \sigma_{z}^{(i)} - J \sum\limits_{i=1}^{N-1} \sigma_{x}^{(i)}\otimes\sigma_{x}^{(i+1)},
\end{equation}
where $g$ is the strength of the transverse field, and $J$ is the
coupling constant for the Ising interaction. \textcolor{black}{As the
  Pauli matrices do not commute with each other, it is impossible to
  minimize the interaction terms and the magnetic field term at the
  same time, meaning that already this simple model lies outside of
  the class of frustration-free Hamiltonians. The transverse field
  Ising} model is known to undergo a quantum phase transition at $g =
J$ from a paramagnetic phase ($g>J$) to a ferromagnetic phase ($g<J$)
\cite{Sachdev1999}. In the following, we will set the energy splitting
of the bath spin $\Delta$ to be identical to the many-body gap $\Delta
E=E_1-E_0$ of the transverse field Ising model, where $E_0$ ($E_1$) is
the energy of the ground state (first excited state). In the
ferromagnetic phase, the ground state becomes doubly degenerate for
large system sizes. Because we are not interested in cooling into a
particular ground state, $E_1$ refers to the first excited state above
the ground state manifold. Below, we will demonstrate that choosing
the bath spin splitting as $\Delta=\Delta E$ leads to optimal cooling,
and we will show how to extract the (a priori unknown) energy gap
$\Delta E$ from the quantum simulation results.


Let us now analyze the cooling performance of the setup by tracking
the system energy $\langle H_{sys}\rangle$ of the transverse field
Ising model in wave-function Monte Carlo simulations of $N=5$ spins,
initially \textcolor{black}{in the experimentally accessible state of all spins} pointing up. Figure~\ref{fig:hsvspar} shows that the
energy of the system decreases rapidly and finally approaches a value
that is close to the numerically calculated ground-state
energy. The cooling performance depends on the choice of
the system-bath coupling $g_{sb}$ and the dissipation rate
$\gamma$. In the following, we assume that the time available for the
cooling remains fixed. Then, if $g_{sb}$ is too small, the cooling
dynamics is very slow. On the other hand, if $g_{sb}$ is too large, then the system and the bath spin will become strongly entangled, and the
cooling performance is reduced. Similarly, if $\gamma$ is too small, then the cooling is slowed down in the same way, while a too large value of
$\gamma$ will lead to a quantum Zeno suppression of the energy
transfer required for the cooling process. Hence, there should be an
optimal choice for $g_{sb}$ and $\gamma$, which leads to a minimum in
energy within the available time.

To find this optimal choice, we \textcolor{black}{use} a model-\textcolor{black}{independent} quantity to
measure the cooling performance. For this, we calculate the fidelity
of the state of the system with respect to the ground-state manifold
of the transverse field Ising model. The fidelity $f$ is given by
\begin{equation}
  f = \langle \Pi_g \rangle = \tr{\rho(t)\Pi_g},
\end{equation}
where $\Pi_g = \sum\limits_i \ketbra{\psi_0^i}$ is the sum of the
projectors onto the ground states \cite{Nielsen2000}. As the inset of
Fig.~\ref{fig:hsvspar}a and \ref{fig:hsvspar}b shows, the ground state
can be prepared with more than 90\% fidelity for the optimal choice of
$g_{sb}=1.15\,g$ and $\gamma=1.9\,g$.

We can also relate the fidelity $f$ to the system energy $\langle H_{sys}\rangle$. For this, we introduce a dimensionless excitation energy $\epsilon$, measured in units of the many-body gap $\Delta E$, i.e.
\begin{equation}
  \epsilon = \frac{\langle H_{sys}\rangle - E_0}{\Delta E}.
\end{equation}
In the low-energy limit $\epsilon \ll 1$ and assuming that the
excitation energy is mostly concentrated in low-energy excitations,
$\epsilon$ is related to the fidelity according to $\epsilon = 1-f$.

We have also checked that our cooling procedure works independently of
the choice of \textcolor{black}{$J/g$, i.e., both in the ferromagnetic phase and in the paramagnet, as well as independently of the initial state (see the Supplementary Materials for details.)} Even in the critical regime ($J/g \sim 1$), where the many-body gap is closing, we observe a similar cooling
performance. To substantiate this point, and also to demonstrate that
our cooling protocol is not limited to a particular model, we turn to the especially challenging case of a critical Heisenberg chain in the following section.


\subsection{Antiferromagnetic Heisenberg model}

As a second paradigmatic quantum many-body model, we investigate the
antiferromagnetic Heisenberg chain, given by the system Hamiltonian
\begin{equation}
  H_{sys} =  J\sum\limits_{i=1}^{N-1} \sum\limits_{j=x,y,z} \sigma_{j}^{(i)}\otimes\sigma_{j}^{(i+1)}.
\end{equation}
This model exhibits an $SU(2)$ symmetry and serves as the critical
point of a Kosterlitz-Thouless transition when the strength of the
$\sigma_z\sigma_z$ interaction is varied \cite{Schollwock2004}. As the
many-body gap vanishes in the thermodynamic limit, this model
represents a particularly challenging case for our cooling
protocol. In addition, the ground state at the critical point is
highly entangled \cite{Latorre2004}; hence, we also test the capability
of our cooling protocol to prepare entangled quantum many-body states.

\begin{figure*}[t]
\begin{tabular}{p{0.1cm}p{8.7cm}p{0.1cm}p{10cm}}
(a) & \vspace{0.05cm} \hspace{-0.1cm} \includegraphics[width=8.5cm]{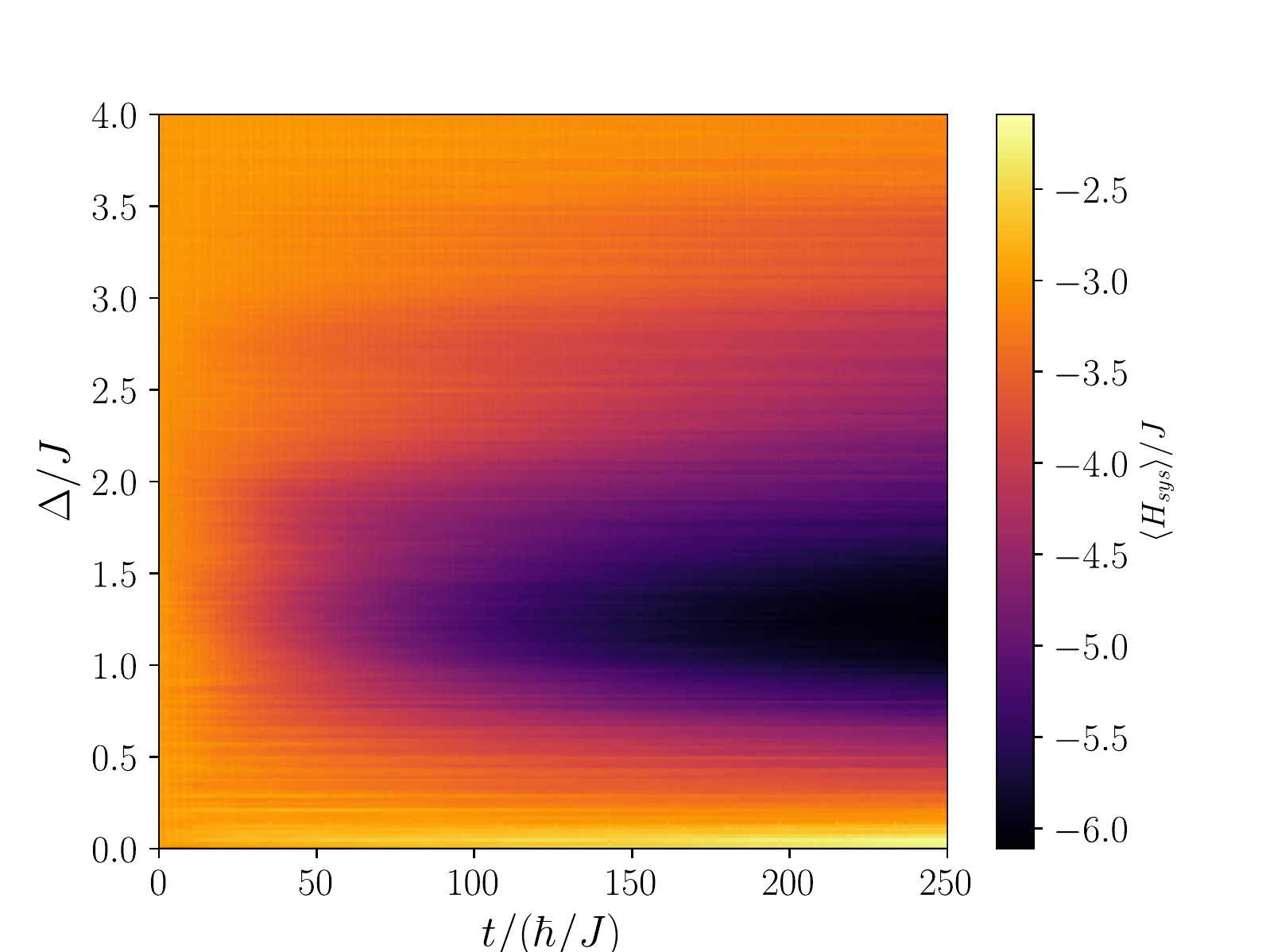} & (b) & \vspace{0.05cm}  \hspace{0.2cm} \includegraphics[width=8.5cm]{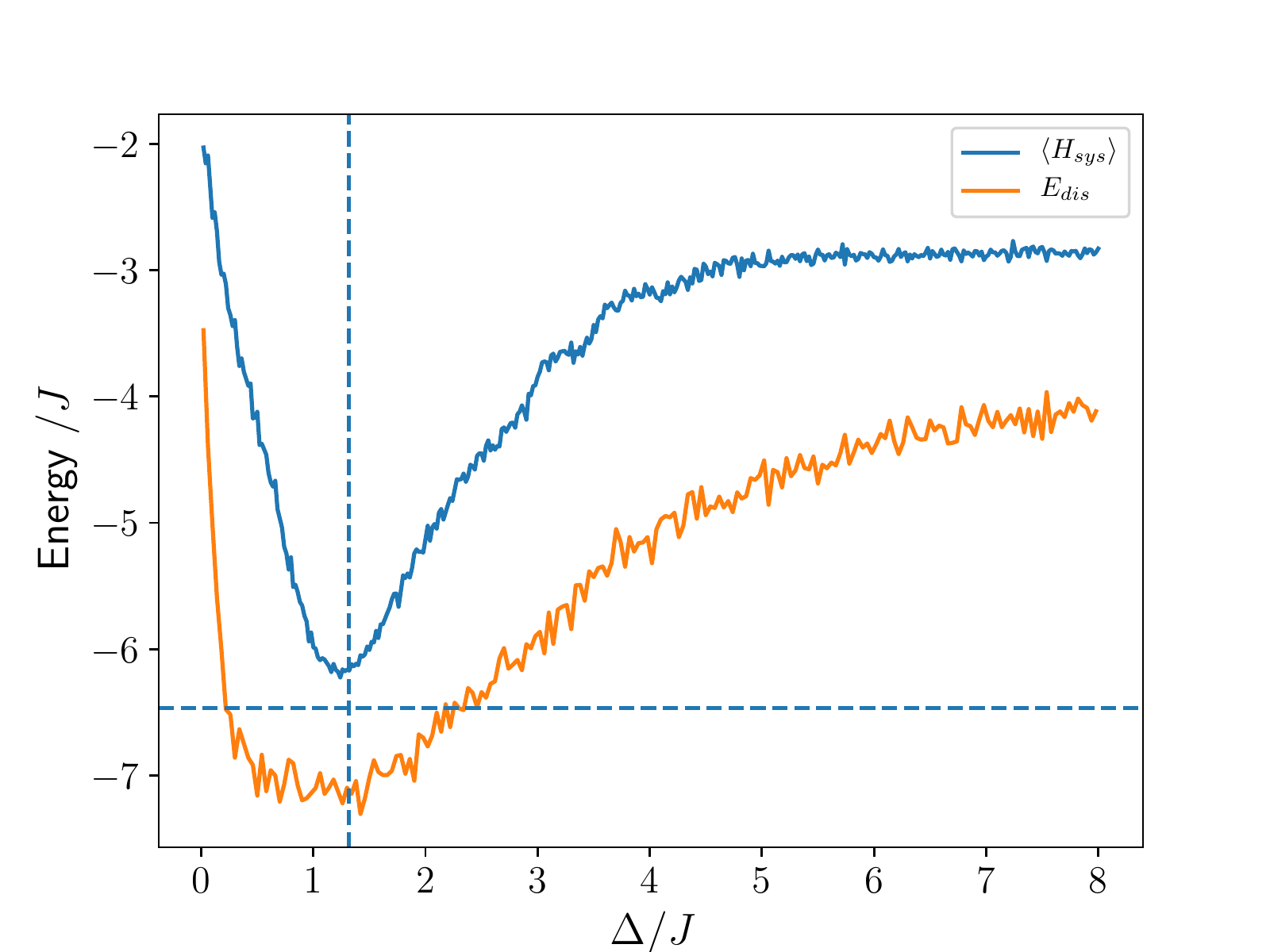}\\
\end{tabular}
\caption{Sympathetic cooling of the antiferromagnetic Heisenberg model
  ($N=4$, \textcolor{black}{$g_{sb}/J = 0.2$, $\gamma/J=0.6$, $f_{x,y,z}=\{ 0.4, 2.3, 0.3\}$}). (a) The efficiency of the cooling procedure depends on the
  choice of the bath spin splitting $\Delta$. (b) The optimal cooling
  leading to the lowest system energy $\langle H_{sys}\rangle$
  corresponds to setting $\Delta$ to the many-body gap $\Delta E$
  (\textcolor{black}{vertical} dashed line). The same minimum is observed when measuring the
  energy $E_{dis}$ that is being dissipated during the cooling
  process. \textcolor{black}{The ground state energy is indicated by the horizontal dashed line.}}

\label{fig:delta}
\end{figure*}

\textcolor{black}{The antiferromagnetic Heisenberg model adds one minor complication concerning its ground state preparation compared with the Ising model. Because of an approximate symmetry conserving certain spin-wave excitations, the ground state cooling performance is limited when the system-bath coupling is restricted to the last spin of the chain. We resolve this issue by an additional system-bath coupling of strength $g_{sb}/2$ to the second last spin of the chain.} Fig.~\ref{fig:delta} shows the cooling performance in terms of the
system energy $\langle H_{sys}\rangle$, \textcolor{black}{with an initial state of spins pointing up and down alternately}, as a function of the splitting
of the bath spin $\Delta$. As in the case of the transverse field
Ising model, $\langle H_{sys}\rangle$ decreases rapidly and reaches a
final value that is close to the ground state energy
$E_0$. In addition, the cooling is optimal when $\Delta$ is chosen to
be identical to the many-body gap $\Delta E$ \textcolor{black}{($f=0.9$)}. Hence,
experimentally measuring $H_{sys}$ as a function of $\Delta$ allows \textcolor{black}{one} to obtain the value of the many-body gap $\Delta E$, which in itself is an important quantity to understand a quantum many-body system. \textcolor{black}{We also find the final state to be highly entangled (see the Supplementary Materials for details).}  

However, on many quantum simulation architectures, it might be
difficult to experimentally measure the system energy $H_{sys}$, as
this will typically require \textcolor{black}{one} to perform tomography on all the operators that appear in the system Hamiltonian. Further challenges arise in architectures where not all coupling constants in the Hamiltonian can be perfectly controlled, leading to additional uncertainties in the estimated value of $\Delta E$.

Fortunately, it is possible to obtain $\Delta E$ by measuring only the
bath spin. The key idea is to measure the energy $E_{dis}$ that is
dissipated during the cooling dynamics. Crucially, this energy is
related to the number of quantum jumps $N_{jump}$ by the relation
$E_{dis} = N_{jump}\Delta$, as a quantum jump will lower the energy of
the bath spin by $\Delta$. We note that there are two different ways
to obtain $N_{jump}$. First, one can directly count the number of
quantum jumps, e.g., by counting the number of emitted photons, if the
dissipative flip of the bath spin is realized by a spontaneous
emission event. In many setups, however, collecting each emitted
photon with high probability might be too challenging. However, as a
second method, one can also obtain $N_{jump}$ via the integrated
probability to find the bath spin in the up state according to
\begin{equation}
  N_{jump} = \gamma\int\limits_0^{t_p}\tr{\sigma_+^{(b)}\sigma_-^{(b)}\rho(t)}\,dt
\end{equation}
where $t_p$ is the total preparation time. As shown in
Fig.~\ref{fig:delta}, the minimum of $E_{dis}$ is almost identical to
the minimum in $H_{sys}$, corresponding to the case where the
splitting of the bath spin $\Delta$ is identical to the many-body gap
$\Delta E$. We note that if the system-bath coupling
  $g_{sb}$ or the dissipation rate $\gamma$ is chosen too large, then the
  difference between the minima in $\langle H_{sys}\rangle$ and
  $E_{dis}$ becomes appreciably larger. We also observe that
$E_{dis}$ is slightly larger in magnitude than the system energy; this
can be attributed to the fact that even in the limit of large times, a
finite probability for quantum jumps remains as the ground state of
the system Hamiltonian is not a perfect dark state of the quantum
master equation \cite{Lemeshko2013a} due to the finite system-bath
coupling $g_{sb}$. \textcolor{black}{These additional jumps can also happen for non optimal values of $\Delta$, leading to a broadening of the dissipated energy $E_{dis}$ in Fig.~3b compared with the system energy $\langle H_{sys}\rangle$.}

\subsection{Efficiency of the cooling protocol}

\begin{figure}[b!]
\includegraphics[width=8.5cm]{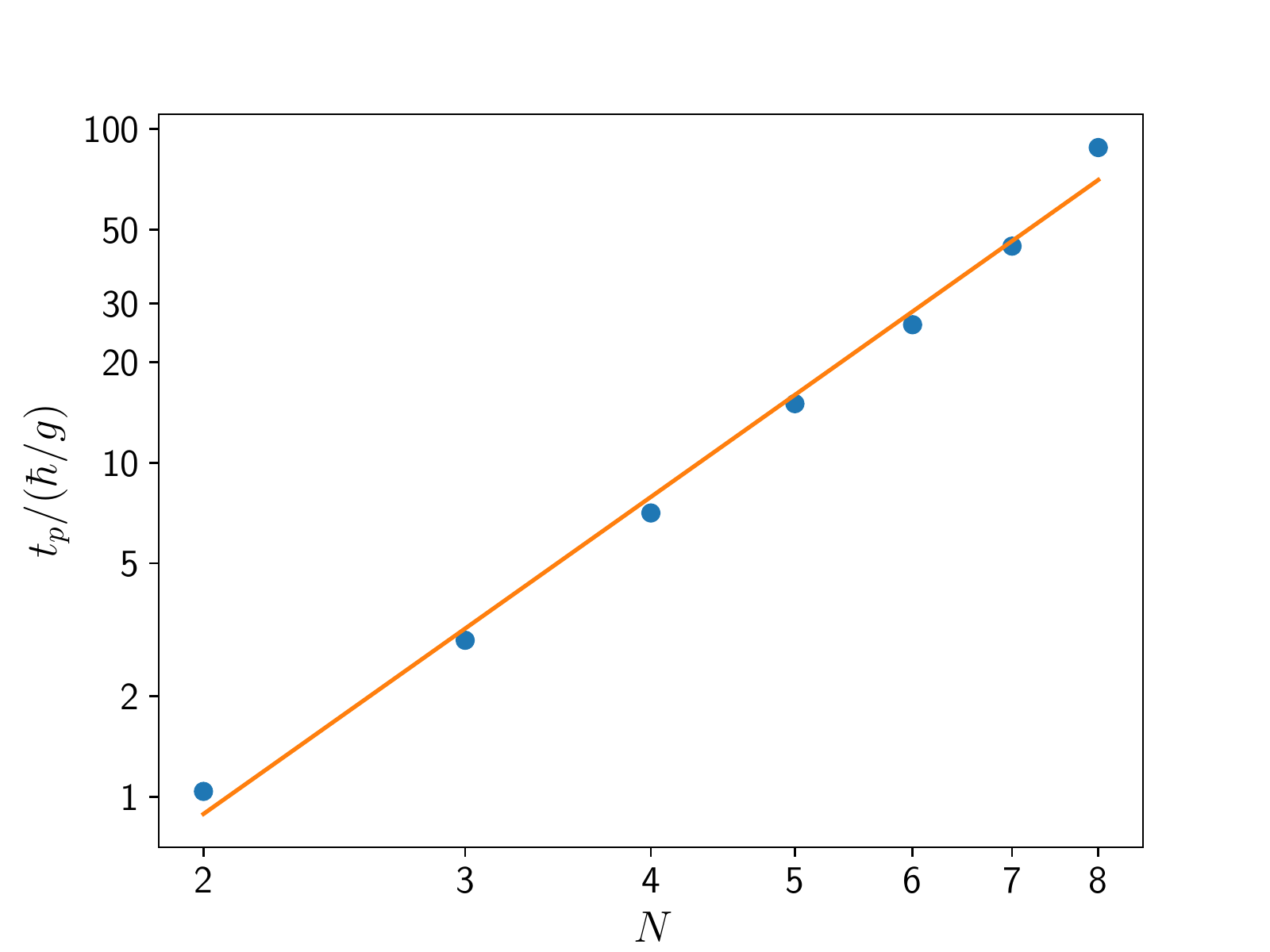}
\caption{Scalability of the cooling protocol. The preparation time
  $t_p$ to reach a final dimensionless energy of $\epsilon = 0.2$
  grows linearly on a log-log scale, i.e., $t_p \sim N^\alpha$. The
  solid line is a fit to the data according to $\alpha = \textcolor{black}{3.1\pm 0.1}$.}
  
\label{fig:scale}
\end{figure}

For any quantum state preparation protocol, it is crucial to determine
how its properties behave when the size of the system is increased. A
protocol is called efficient when the resources required
(i.e., the preparation time) grow at most polynomially with the system
size. To determine the scaling with system size in an
unbiased way, we compute the preparation time $t_p$ that is required
to cool the system down to a fixed dimensionless energy $\epsilon$,
while the system bath coupling $g_{sb}$ and the dissipation rate
$\gamma$ are chosen such that the cooling is optimal. \textcolor{black}{Within our numerical simulations, we use a standard nonlinear optimization scheme (see the Methods section for details). In an actual quantum simulator, one can use a hybrid algorithm in which the energies measured on the quantum device are fed back into the classical optimization algorithm \cite{Kokail2019}.}

Figure \ref{fig:scale} shows the scaling behavior of $t_p$ for the
transverse field Ising model. Although the system is cooled across the
phase transition into the ferromagnet, the preparation time grows only
polynomially with the system size. \textcolor{black}{The same scaling is
  also observed for the antiferromagnetic Heisenberg model (see the
  Supplementary Materials for details).} This scaling behavior underlines
that our cooling procedure is already scalable when using only a
single bath spin. As the number of particles is often a scarce
resource in a quantum simulator, the required minimal overhead for the
initialization allows \textcolor{black}{us} to use almost all of the
particles for the actual quantum simulation.

\subsection{Performance under decoherence}

So far, the only source of decoherence in our considerations stems
from the dissipative flips of the bath spin. However, in most quantum
simulation architectures, there will also be unwanted decoherence
processes in the system performing the quantum simulation. Therefore,
it is crucial to determine the consequences of this additional
decoherence on the performance of our cooling protocol.

As an additional source of decoherence, we consider $\sigma_z$ spin
flips in the quantum simulation of the transverse field Ising model,
applied with a rate $\kappa$ to all $N$ spins of the quantum
simulator. In the ferromagnetic phase, such a spin flip will create
two neighboring domain-wall excitations, i.e., when applied to the
ground state, the dimensionless energy will approximately increase to
$\varepsilon \approx 2$. This type of decoherence represents a worst
case scenario of all local decoherence
processes. Hence, we expect that this scenario is
  quite generic and that our findings should also apply to other
  many-body models.

To analyze the consequences of these additional decoherence
channels, we consider the quantity $\kappa t_p$, which is essentially
the probability of any spin to undergo a decoherence event during the
preparation time. Then, tracking how the energy $\epsilon$ behaves as
a function of $\kappa t_p$ allows us to assess the robustness of our
cooling protocol under additional decoherence.

\begin{figure}
\includegraphics[width=8.5cm]{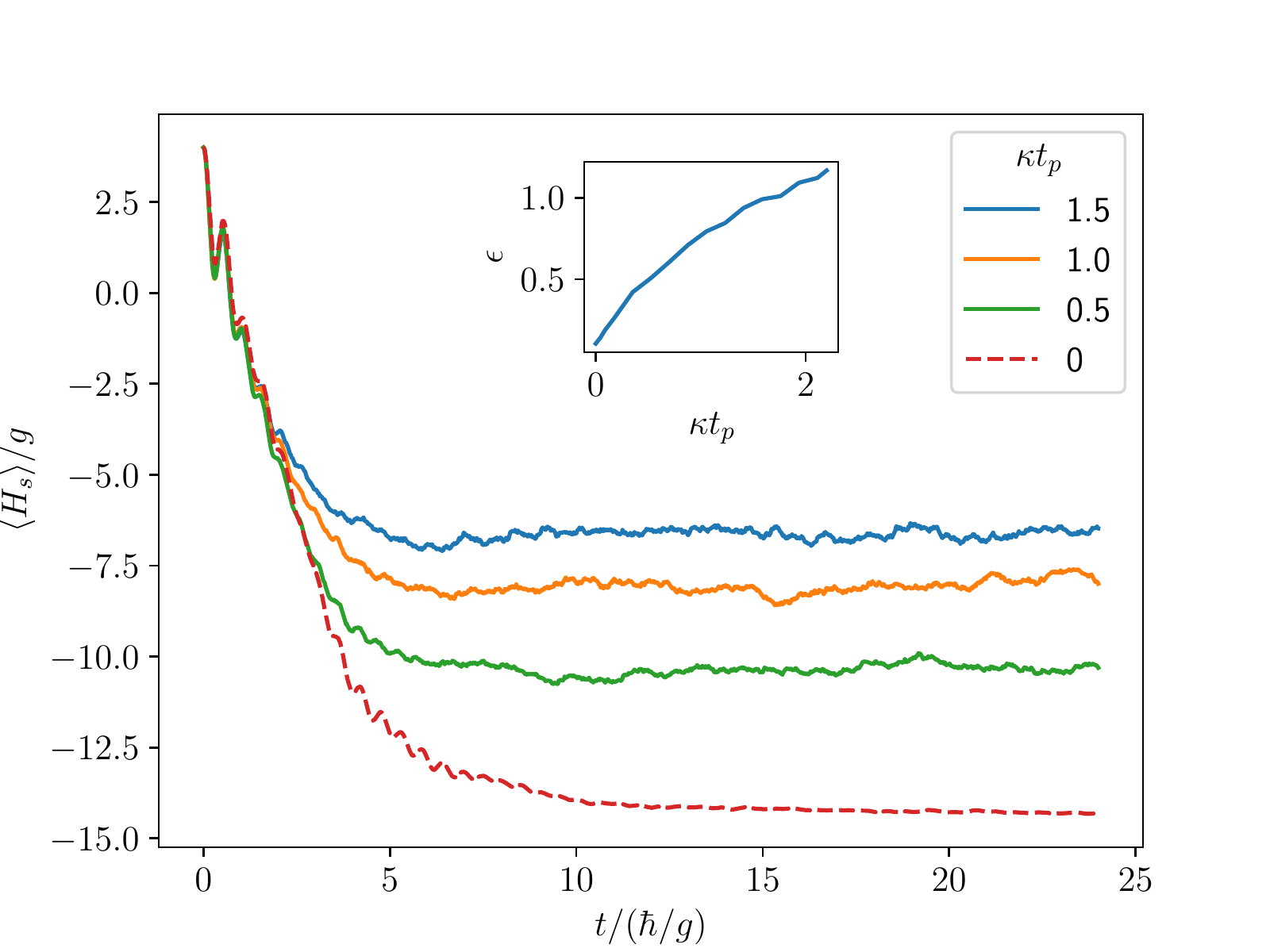} 
\caption{Cooling performance in the presence of decoherence in the
  quantum simulator for the transverse field Ising chain ($J/g = 5$,
  $N=4$). The inset shows the dimensionless energy $\epsilon$ as a
  function of the product $\kappa t_p$, where $t_p$ was
  taken from the dynamics without decoherence corresponding to a
  ground state preparation fidelity of $f=0.9$ (dashed line).}

\label{fig:noise}
\end{figure}

Figure \ref{fig:noise} shows the system energy for different
decoherence rates, from which the behavior of $\epsilon$ is
calculated. Crucially, we find that the system contains one
excitation, $\epsilon \approx 1$ at a value of $\kappa t_p \approx
2$. This means that the system picks up one excitation when on average all the spins have undergone a decoherence event. This is in
stark contrast to the scaling observed in adiabatic state
preparation protocols, where the error probability is typically given
by the probability that a single spin undergoes a decoherence
event, i.e., proportional to $N\kappa t_p$ \cite{Weimer2012}. This
improved robustness against decoherence can be attributed to the fact
that our state preparation protocol itself is dissipative and
therefore can self-correct decoherence events.

\subsection{Experimental realization}

The proposed initialization protocol can be implemented in a trapped ion system with state-of-the-art technology, e.g. by confining a 1D ion string in a linear Paul trap.
Here, we propose an implementation with $^{40}$Ca$^+$-ions in a setup similar to the one described in \cite{Jurcevic2014}.
The spin states are encoded in the optical qubit, $\ket{\!\downarrow}=\ket{S_{1/2},m=+1/2}$ and $\ket{\!\uparrow}=\ket{D_{5/2},m=+5/2}$ (see Fig.~\ref{fig:setup}c) with an energy splitting of $\hbar \omega_0$, coherently manipulated by radial laser beams.
E.g. the rightmost ion serves as the bath spin (index $b$), while its laser-induced coupling to the neighboring ion (index $s$) implements the system-bath coupling. The bath ion can be isolated from the system interaction by shelving the population to an auxiliary state $\ket{\text{aux}}_b=\ket{D_{5/2},m=-5/2}_b$ with a laser beam addressing only the bath ion. An experimental realization requires the implementation of the system and system-bath Hamiltonians. For simplicity, we suggest to implement $H_{sys}$ and $H_{sb}$ in an interleaved fashion by trotterizing the total interaction~\cite{Lloyd1996,Lanyon2011}.

In trapped ion systems, $H_{sys}$ for the transverse field Ising model~\cite{Friedenauer2008} has been realized with up to 53 qubits~\cite{Zhang2017}. For this purpose, a global bichromatic laser beam with frequency $\omega_0\pm\delta$ implements a gate operation by coupling to all radial modes. If $\delta$ is larger than the center-of-mass mode frequency, then the resulting spin-spin coupling coefficient shows a power law scaling $J_{i,j}\propto 1/|i-j|^\alpha$~\cite{Kim2009}, where $\alpha$ can be varied between $0$ and $3$ by changing the radial confinement. Implementation of the Heisenberg model is possible by interleaving the spin-spin coupling gates with single-qubit rotations performing a basis change from $\sigma_x$ to $\sigma_y$ and $\sigma_z$.

We propose to implement $H_{sb}$ with a separate laser that provides single ion addressing for the bath spin and the neighboring system spin. A M\o lmer-S\o rensen gate \cite{Benhelm2008,Roos2008} on the radial motional modes bridges two different energy gaps, $\omega_s$ and $\omega_0$, similar to a two-species gate~\cite{Tan2015}, and provides a $\sigma^{(N)}_x\sigma^{(b)}_x$-type coupling of the spins. For the bath spin, the laser frequencies will be  $\omega_0\pm\delta$ and for the system spin will be $\omega_s\pm\delta$ with $\omega_s=\Delta E/\hbar$ for optimal cooling. Tuning the latter frequency corresponds to searching for the resonance condition described in the main text. Again, $\sigma^{(N)}_x\sigma^{(b)}_x$-gates interleaved with single qubit rotations on both ions implement $\sigma^{(N)}_x\sigma^{(b)}_x$, $\sigma^{(N)}_y\sigma^{(b)}_y$, and $\sigma^{(N)}_z\sigma^{(b)}_z$. \textcolor{black}{The coupling between the bath spin and the second last spin for the Heisenberg model can be realized by extending the addressing laser to the second last spin such that the power law of the system-bath interaction has an exponent $\alpha_{sb} \sim 1$. This comes at the expense of an additional interaction between the last and the second-last spins of the system, which is significantly weaker than $J$ and can therefore be neglected. This additional coupling may also be canceled using an additional addressing laser.}

Assuming $\Delta E$ is already known, repumping from $\ket{\!\uparrow}_b$ to P$_{3/2}$ and a subsequent spontaneous decay to $\ket{\!\downarrow}_b$ on the bath ion can be used to provide a channel for dissipation. The strength of dissipation, $\gamma$, within the trotterized scheme can be adjusted by the repumping laser intensity, i.e. the repumping probability during each Trotter cycle. For determination of $\Delta E$ by recording $N_{jump}$, every scattered photon during the repump process has to be detected. This is accomplished by an electron shelving scheme in which the population in $\ket{\!\downarrow}_b$ is hidden in state $\ket{\text{aux}}_b$ and a potentially scattered photon bringing the bath ion from $\ket{\!\uparrow}_b$ to $\ket{\!\downarrow}_b$ is detected by measuring fluorescence on the $\ket{\!\downarrow}_b$ (S$_{1/2}$) to P$_{1/2}$ transition. To avoid a perturbation of the system spins, the detection laser has to be tightly focused onto the bath ion.

\begin{figure}[b!]
\includegraphics[width=8.5cm]{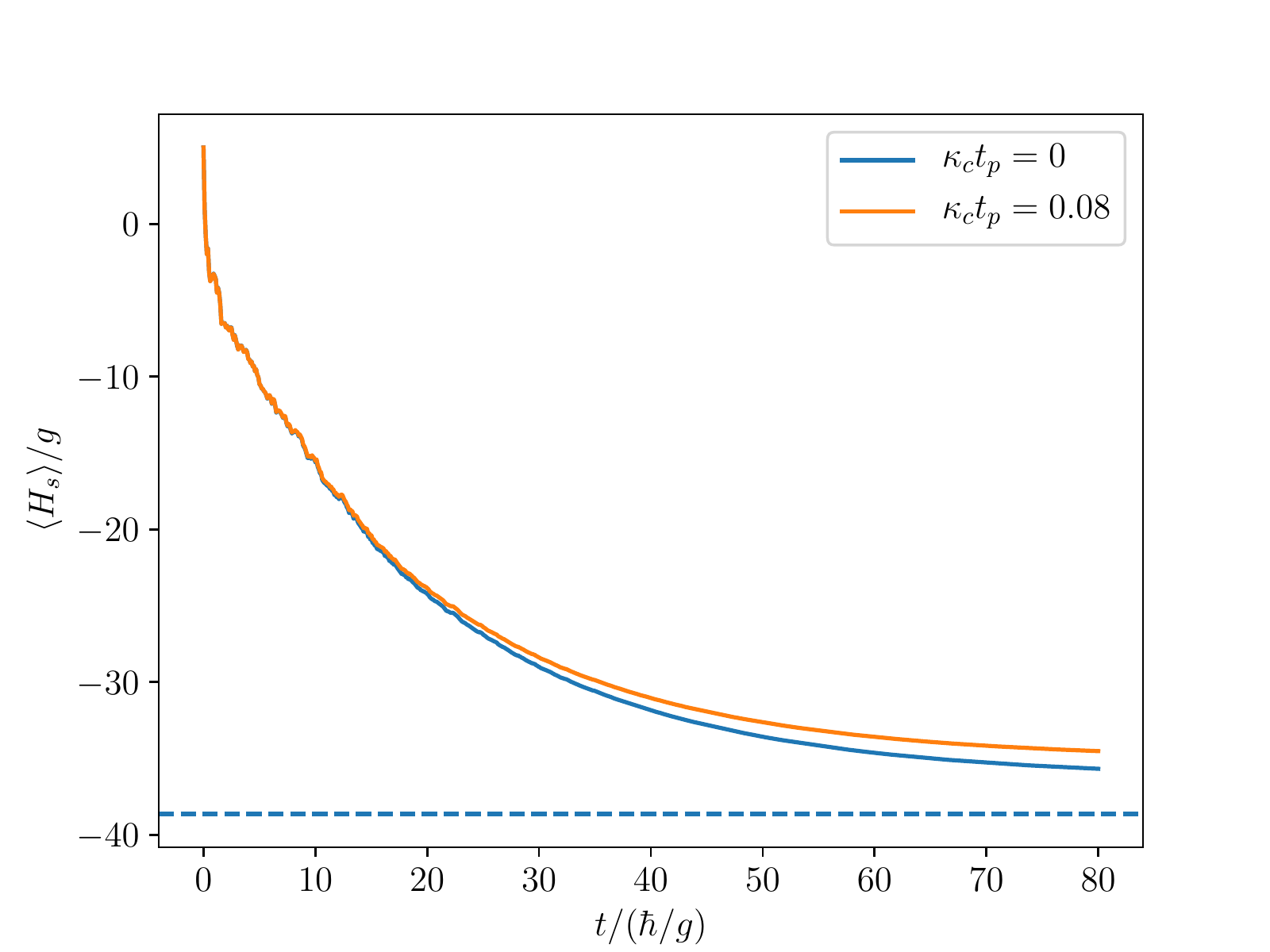}
\caption{\textcolor{black}{Cooling performance of an Ising-like chain of $5+1$ ions of $t_p=80\hbar/g = 24$~ms. The blue line shows the dynamics in the decoherence-free case resulting in a fidelity of $f=0.92$, while the orange line indicates the dynamics under a collective decoherence mechamism with rate $\kappa_c = 3.3$~Hz, resulting in $f=0.89$. The dashed line indicates the ground state energy of the system.}}
\label{fig:lr}
\end{figure}

To be more specific, we assume \textcolor{black}{5} $^{40}$Ca$^+$ ions in a linear chain with single ion axial and radial trapping frequencies of $\omega_z=2\pi\times 0.15$~MHz and \textcolor{black}{$\omega_r=2\pi\times 0.5$~MHz}, respectively \cite{Jurcevic2014}. With a resonant Rabi frequency of $2\pi\times 125$~kHz for all ions, $J_{i,j}$ ranges between \textcolor{black}{$2\pi\times 2.7$~kHz and $2\pi\times 1.2$~kHz} for a detuning of \textcolor{black}{$\delta-\omega_r\approx 2\pi\times 10.5$~kHz, while the largest Lamb-Dicke parameter is given by $\eta_{\textrm{max}} =0.128$.}
For these parameters, the spacing between the bath spin and the nearest system spin of around \textcolor{black}{$14~\upmu \text{m}$} is sufficiently large to provide a factor of $10^{-7}$ suppression of the scattering rate for the electron shelving detection on the neighboring ion for a beam focused to $2.6~\upmu \text{m}$ on the bath ion.

\textcolor{black}{In such a setup, the dominant decoherence mechanism is arising from global magnetic field fluctuations. This process can be expressed in terms of a jump operator of the form $c' = \sqrt{\kappa_c}\sum_i\sigma_z^{(i)}$, assuming a decoherence rate of $\kappa_c = 3.3$~Hz \cite{Ruster2016}.} \textcolor{black}{Fig.~\ref{fig:lr} shows the cooling of such a system to an optimized ground state fidelity of $f=0.92$ in the decoherence-free case, while the presence of decoherence leads to a fidelity of $f = 0.89$.}

An alternative to single ion addressing is to employ another isotope for the bath ion, such as $^{44}$Ca$^+$. The large isotope shifts of 850~MHz on the S$_{1/2}$-P$_{1/2}$ transition and 5.3~GHz on the qubit transition \cite{Solaro2018, Gebert2015} will significantly relax the focussing requirements at the expense of having to achieve an appropriately ordered ion crystal \cite{Splatt2009}.

\section{Discussion}

Here, we demonstrated how adding a dissipatively driven auxiliary
particle can sympathetically cool a quantum simulator into low-energy
states. Our approach is efficient even when using only a single
bath spin, and it exhibits strong robustness against unwanted
decoherence occurring in the quantum simulator. Future directions
include investigating the scaling behavior when optimally varying the
coupling constants of the bath in time and when adding multiple
bath spins. In the latter case, it will also be of interest to choose
different splittings of the bath spins, allowing engineering of tailored
bath spectral functions for the quantum simulator.

\section{Materials and Methods}

\subsection{Numerical simulations}

All numerical simulations were performed using a wave-function Monte
Carlo approach provided by the QuTiP library \cite{Johansson2013},
extended to a massively parallelized version
\cite{Raghunandan2018}. Results were obtained by averaging over 1,000
Monte Carlo trajectories. We note that we are interested in the long
time limit of a weakly dissipative system, i.e., a regime where tensor
network algorithms are breaking down
\cite{Kshetrimayum2017}. Numerical optimization of the coupling
constants $g_{sb}$ and $\gamma$ was carried out using a Nelder-Mead
algorithm. \textcolor{black}{We typically obtain convergence within
  approximately 50 runs of the simulation, which does not
  significantly depend on the size of the system.}

\section{Acknowledgements}

  This work was funded by the Volkswagen Foundation and the DFG within SFB 1227 (DQ-mat, projects A01, A04, and B05).

  \section{Author contributions}

  M.R. and H.W. designed the cooling protocol. M.R. performed the
numerical simulations. F.W., C.O., and P.O.S. designed the experimental
implementation proposal. All authors contributed to the writing of the manuscript.

  \section{Competing interests}

  The authors declare that they have no competing interests.

  \section{Data and materials availability}

  All data needed to evaluate the conclusions in the paper are present
  in the paper and/or the Supplementary Materials. Additional data
  related to this paper may be requested from the authors.


\end{document}



%
%

\title{Supplementary Material for ``Initialization of quantum simulators by sympathetic cooling''}

\author{Meghana Raghunandan}
\email{meghana.raghunandan@itp.uni-hannover.de}
\affiliation{Institut f\"ur Theoretische Physik, Leibniz Universit\"at Hannover, Appelstra{\ss}e 2, 30167 Hannover, Germany}
\author{Fabian Wolf}
\affiliation{QUEST Institut, Physikalisch-Technische Bundesanstalt, 38116 Braunschweig, Germany}
\author{Christian Ospelkaus}
\affiliation{QUEST Institut, Physikalisch-Technische Bundesanstalt, 38116 Braunschweig, Germany}
\affiliation{Institut f\"ur Quantenoptik, Leibniz Universit\"at Hannover, Welfengarten 1, 30167 Hannover, Germany}
\author{Piet O. Schmidt}
\affiliation{QUEST Institut, Physikalisch-Technische Bundesanstalt, 38116 Braunschweig, Germany}
\affiliation{Institut f\"ur Quantenoptik, Leibniz Universit\"at Hannover, Welfengarten 1, 30167 Hannover, Germany}
\author{Hendrik Weimer}
\affiliation{Institut f\"ur Theoretische Physik, Leibniz Universit\"at Hannover, Appelstra{\ss}e 2, 30167 Hannover, Germany}

\maketitle

\section{Energy-level representation of the cooling protocol}

\begin{figure*}[h]
\begin{tabular}{p{0.1cm}p{8.7cm}p{0.1cm}p{10cm}}
(a) & \vspace{0.05cm} \hspace{-0.1cm} \includegraphics[width=8.5cm]{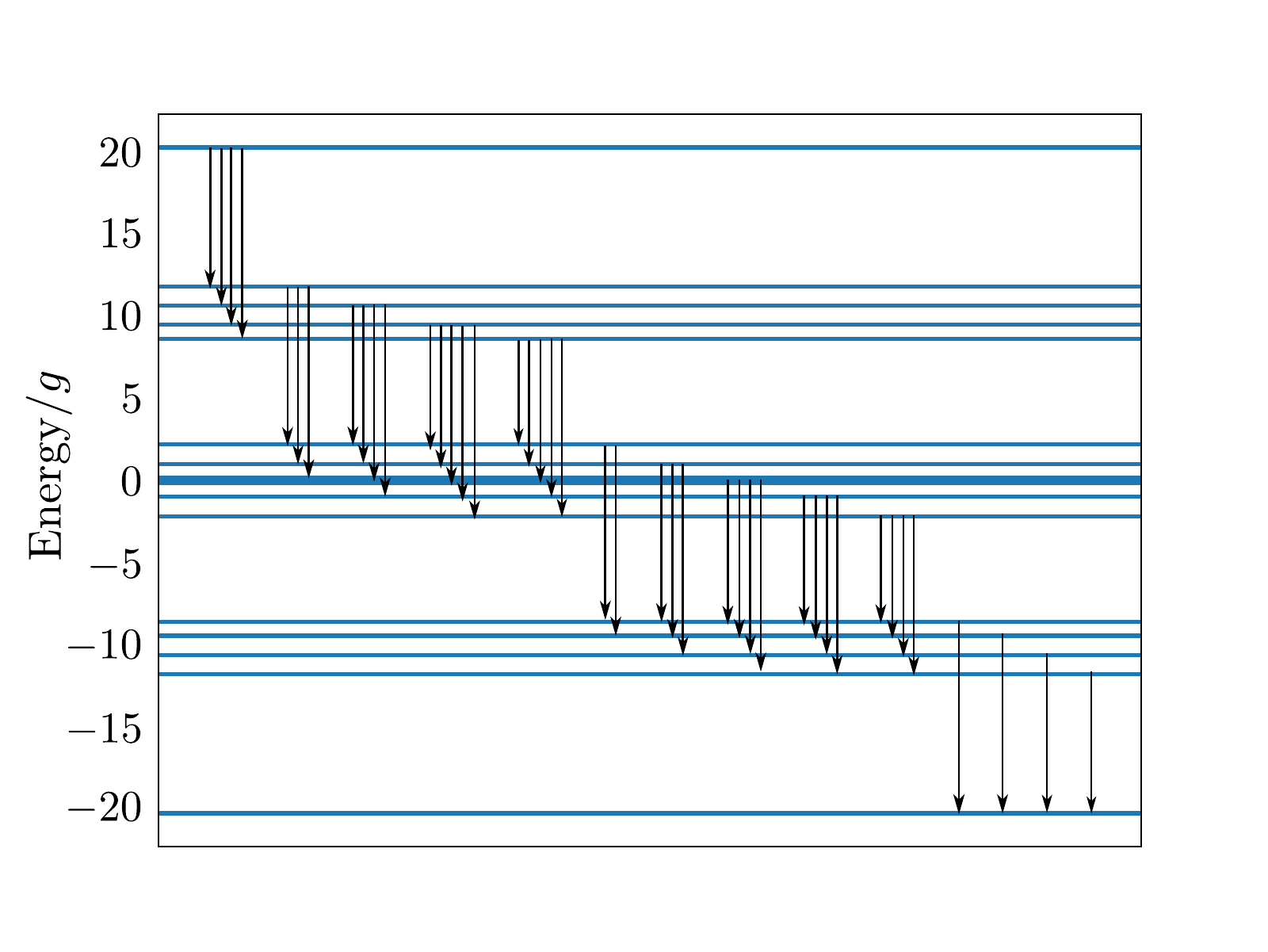} & (b) & \vspace{0.05cm}  \hspace{0.2cm} \includegraphics[width=8.5cm]{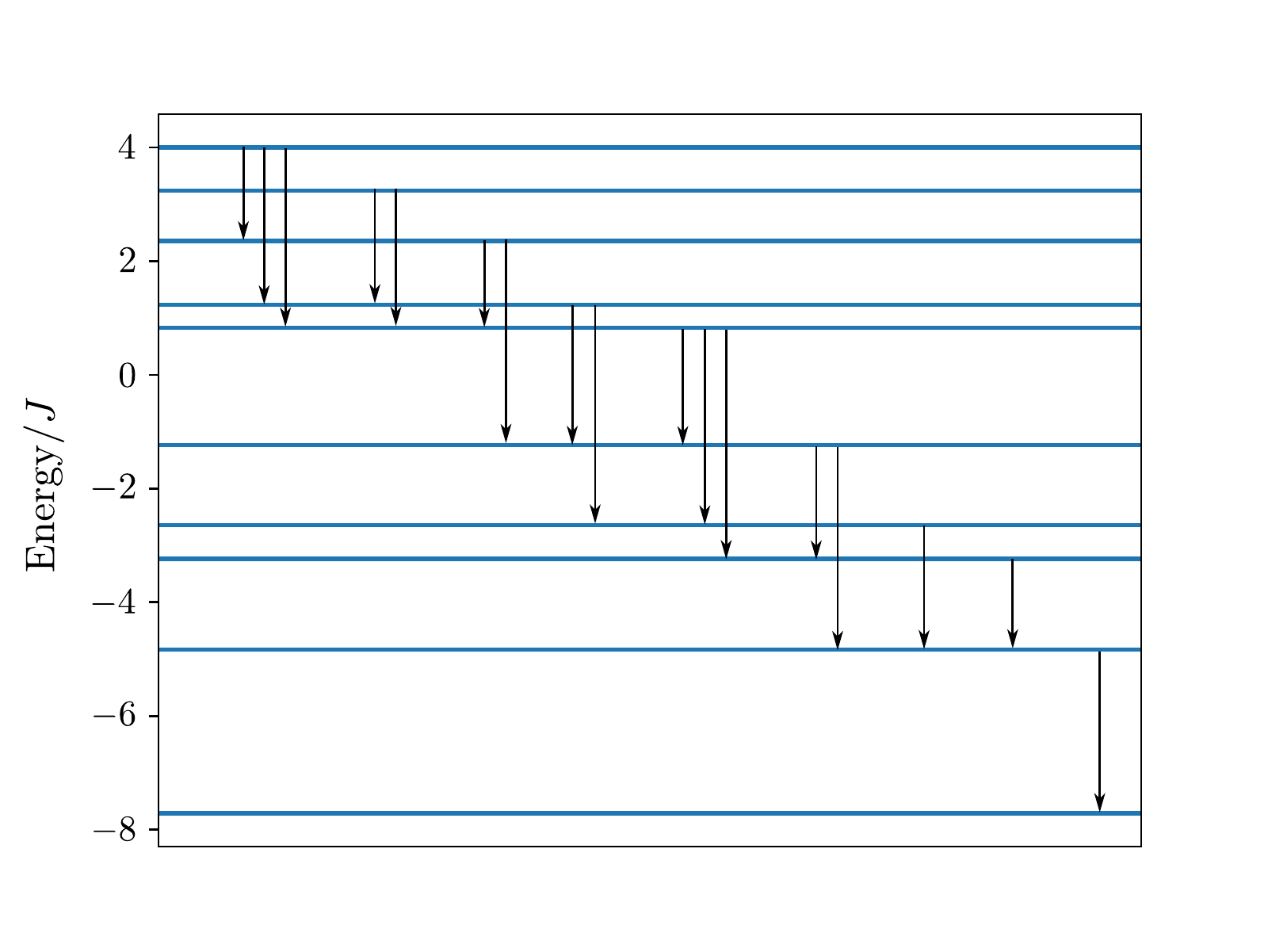}\\
\end{tabular}

\caption{Possible paths via which an excitation can be cooled down to the ground state. Each black arrow corresponds to an energy difference $\Delta - \gamma \leq E_i-E_j \leq \Delta +\gamma $. Each cooling step leads to a reduction of the energy of the system, eventually reaching the ground state.  The energy levels are shown for (a) the Ising model ($N=5$, $J/g=5$, $\gamma/g = 3.5$) and (b) the Heisenberg model ($N=5$, $\gamma/J=1.26$).}
\label{fig:energies}
\end{figure*}

The physical mechanism behind our cooling protocol can be understood
by looking at the transitions between the energy levels $E_i$ of the
many-body Hamiltonian $H_{sys}$. In a very simplified picture, we
assume that a cooling transition is possible when its energy
difference $E_i-E_j$ is within the energy window provided by the bath
spin, whose splitting $\Delta$ is broadened by the decay rate
$\gamma$. Fig.~\ref{fig:energies} shows all possible transitions
between energy levels with energy differences of $\Delta \pm \gamma$. After the energy is transferred from the many-body system to the bath spin, it can be dissipatively removed. After many such processes, the system is cooled down to its ground state. As seen from Fig.~\ref{fig:energies}, there are multiple possible paths for cooling of the excited states, which enables high-fidelity ground state preparation. \textcolor{black}{Note that the system-bath interaction explicitly breaks any possible symmetry of the system Hamiltonian, therefore also symmetry-breaking cooling transitions are possible.}

\section{Cooling in the paramagnetic and the critical regime of the Ising model}

\begin{figure*}[h!]
\begin{tabular}{p{0.1cm}p{8.7cm}p{0.1cm}p{10cm}}
(a) & \vspace{0.05cm} \hspace{-0.1cm} \includegraphics[width=8.5cm]{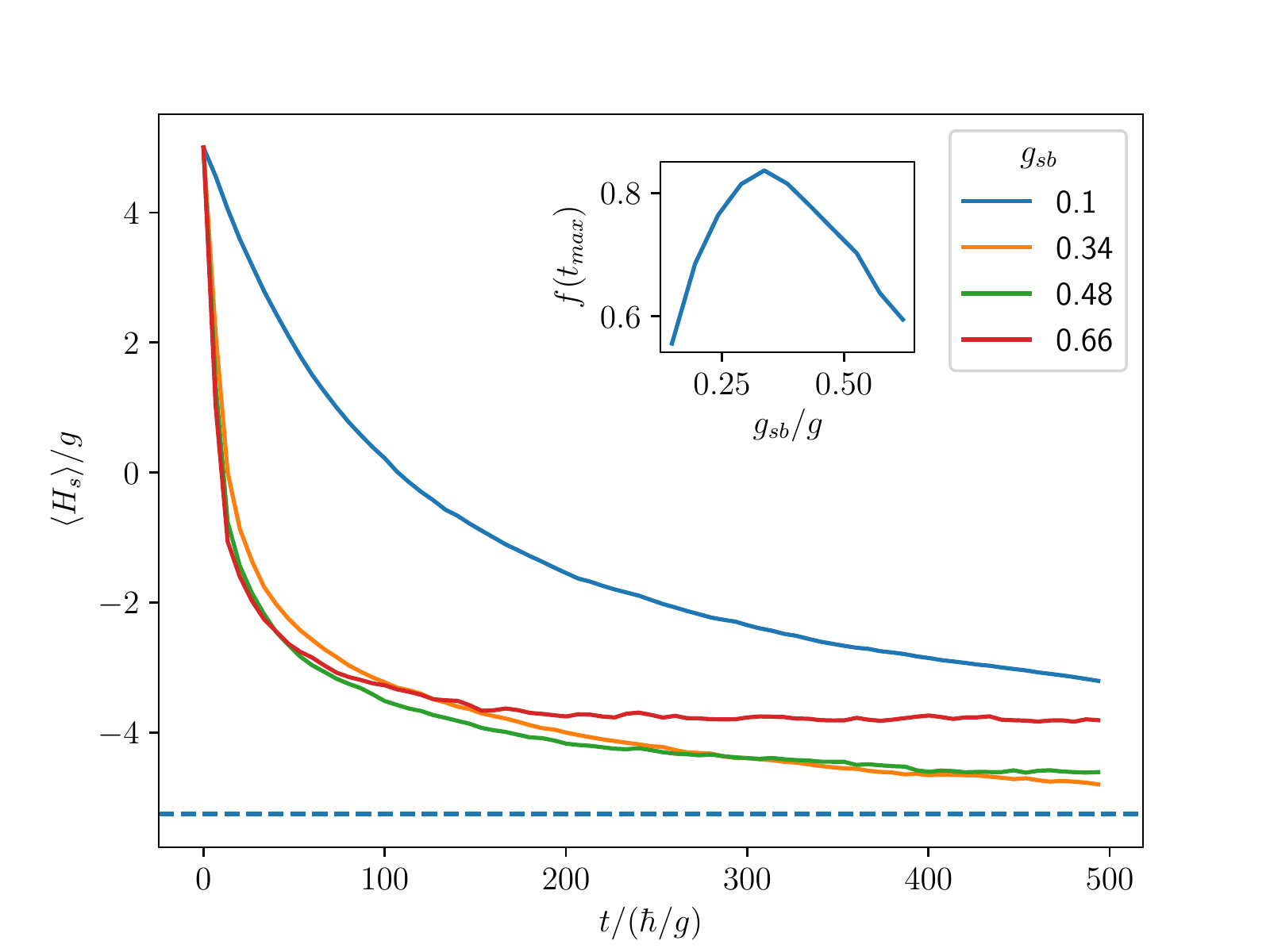} & (b) & \vspace{0.05cm}  \hspace{0.2cm} \includegraphics[width=8.5cm]{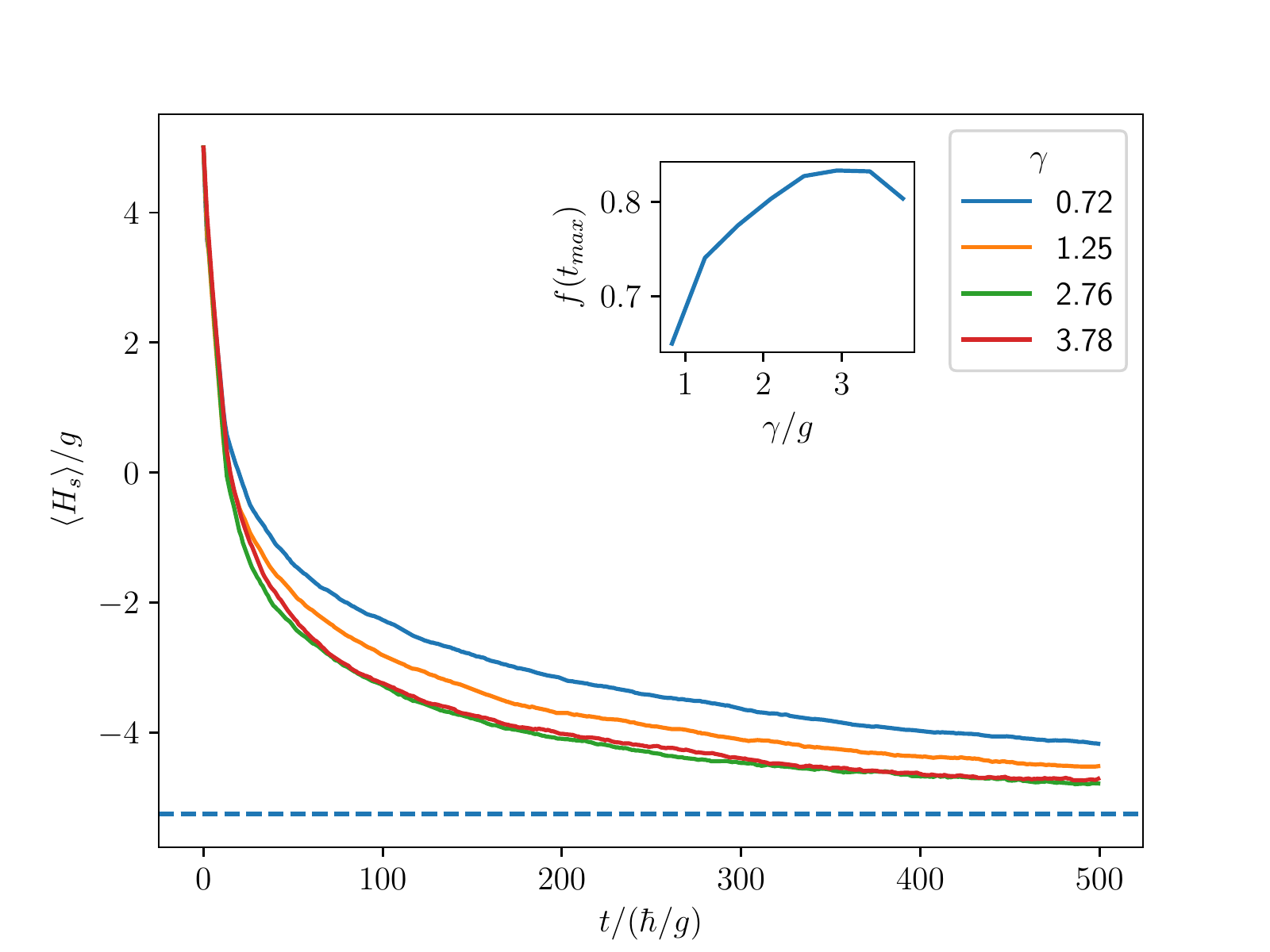}\\

(c) & \vspace{0.05cm} \hspace{-0.1cm} \includegraphics[width=8.5cm]{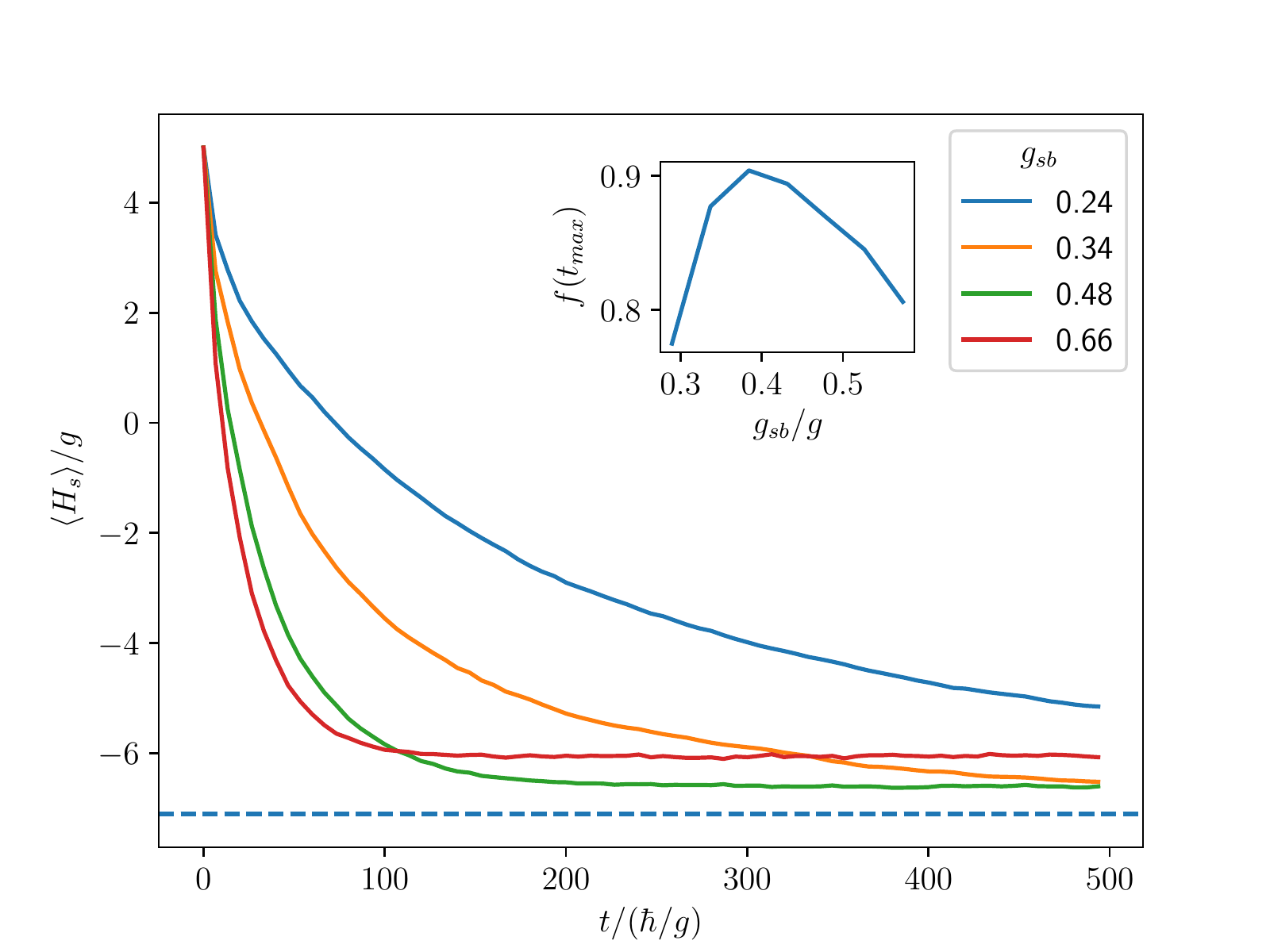} & (d) & \vspace{0.05cm}  \hspace{0.2cm} \includegraphics[width=8.5cm]{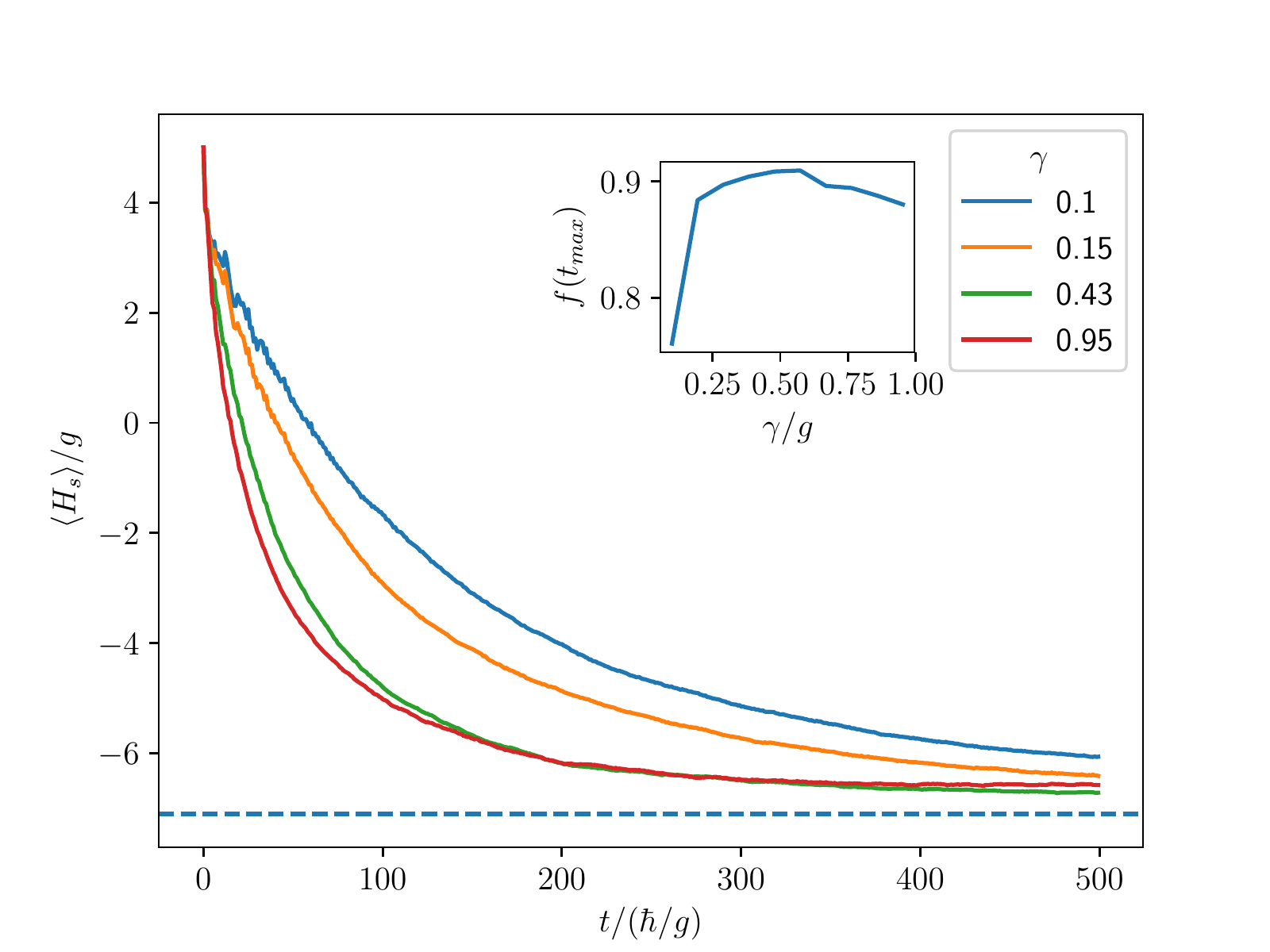}\\
\end{tabular}
\caption{Sympathetic cooling of the transverse Ising model ($N=5$, $f_{x,y,z}=\{ 1, 1.1, 0.9 \}$) in the paramagnetic phase ($J/g=0.2$) (a-b) and in the critical regime ($J/g = 1.4$) (c-d).}
\label{fig:regimes}
\end{figure*}

Here, we check the performance of our cooling protocol in the paramagnetic ($g \gg J$) and the critical ($g \sim J$) regimes. Fig.~\ref{fig:regimes} (a) and (b) show the cooling of an Ising chain with $N=5$ spins in the paramagnetic phase whereas Fig~\ref{fig:regimes} (c) and (d) show the cooling in the critical regime. We observe the existence of optimal values of the parameters, $g_{sb}$ and $\gamma$ for maximum cooling similar to the case of the ferromagnetic Ising model. Note that due to finite size scaling, the critical regime for a system of $N=5$ spins is not at $J/g =1$ but rather at $J/g=1.4$ which we have determined using the peak of the magnetic susceptibility.

\section{Dependence of the cooling performance on the initial state}

\begin{figure*}[h]
\includegraphics[width=8.5cm]{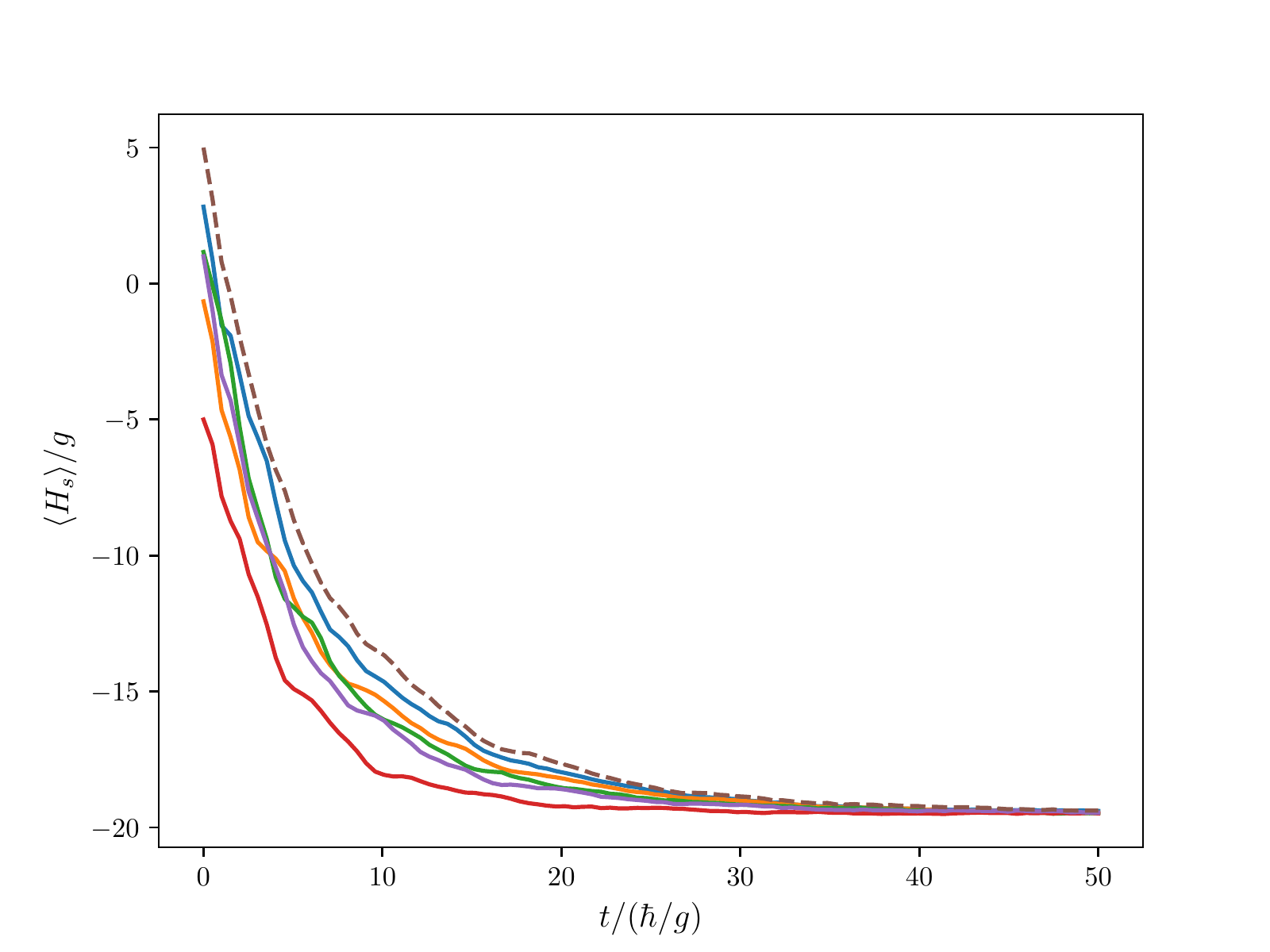}
\caption{Cooling performance of the transverse field Ising model in
  the ferromagnetic phase for various initial states. The dashed line
  corresponds to the case of all spins initially pointing up ($N=5$,
  $J/g = 5$, $g_{sb}/g=1.15$, $\gamma/g=1.9$, $f_{x,y,z}=\{ 1, 1.1,
  0.9 \}$).}
\label{fig:init}
\end{figure*}

While it is convenient to use an experimentally accessible state as
the initial state for the cooling protocol, one may ask whether the
cooling performance depends on the choice of the initial state. We
investigate this dependence by choosing several product states with
random configurations of up and down spins. In Fig.~\ref{fig:init}, we
show the dynamics of the cooling for the transverse field Ising model. We
see that the initial energies differ significantly, while the overall
dynamics remains very similar. This picture is consistent with the
cooling timescale being a property of the combined system-bath
Liouvillian that does not depend on the state of the system.

\clearpage

\section{Efficiency of the cooling protocol for the Heisenberg model}

In the same way as for the Ising model, we determine the scaling of
the preparation time $t_p$ with the system size $N$ for the
antiferromagnetic Heisenberg model. Within spin-wave theory
\cite{Kubo1952}, one can see that the model exhibits a particular
symmetry that makes cooling slightly more challenging than for the
Ising model. The reason is a parity symmetry that arises when
partitioning the model into two sublattices when constructing the
spin-wave theory. Hence, it is possible that some excitations cannot
be cooled when the bath spin is coupled to only the last site (and
hence only to one of the two sublattices). Coupling the bath spin to
the second-last site as well (here, we choose a coupling strength of
$g_{sb}/2$) resolves the problem, i.e., $H_{int} = g_{sb}
\sum\limits_{x,y,z} f_i \sigma_{i}^{(N)}\sigma_{i}^{(b)} + \frac{g_{sb}}{2}\sum\limits_{x,y,z} f_i \sigma_{i}^{(N-1)}\sigma_{i}^{(b)}$.

\begin{figure*}[h!]
\includegraphics[width=8.5cm]{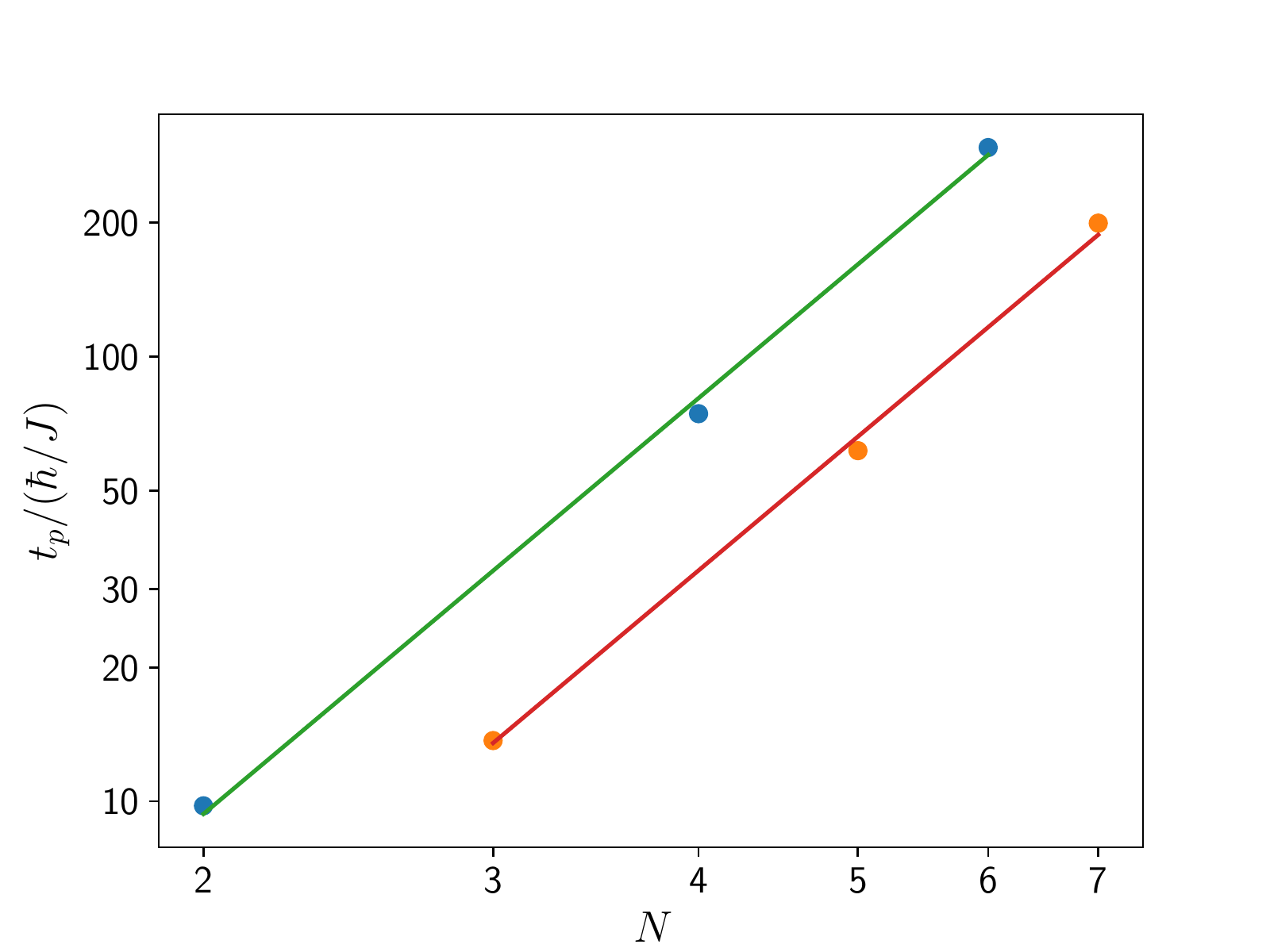}
\caption{Scalability of the protocol for the antiferromagnetic Heisenberg model. The preparation time $t_p$ to reach a final dimensionless energy of $\epsilon = 0.2$ grows linearly on a log-log scale, i.e., $t_p \sim N^\alpha$ as in the case of transverse Ising model. The green \textcolor{black}{($N$ even)} and red \textcolor{black}{($N$ odd)} solid lines are the fits to the data with a common exponent $\alpha$ according to $\alpha = 3.11 \pm 0.01$.}
\label{fig:hei_scale}
\end{figure*}

Fig. \ref{fig:hei_scale} shows that the optimal preparation time $t_p$
for the antiferromagnetic Heisenberg model scales polynomially with
the system size $N$. The smaller preparation times for odd system
sizes can be attributed to the fact that their ground states are
doubly degenerate which provides for more pathways for faster ground
state preparation.

\section{Entanglement measure for the ground state cooling of the Heisenberg model}

\begin{figure*}[h!]
\includegraphics[width=8.5cm]{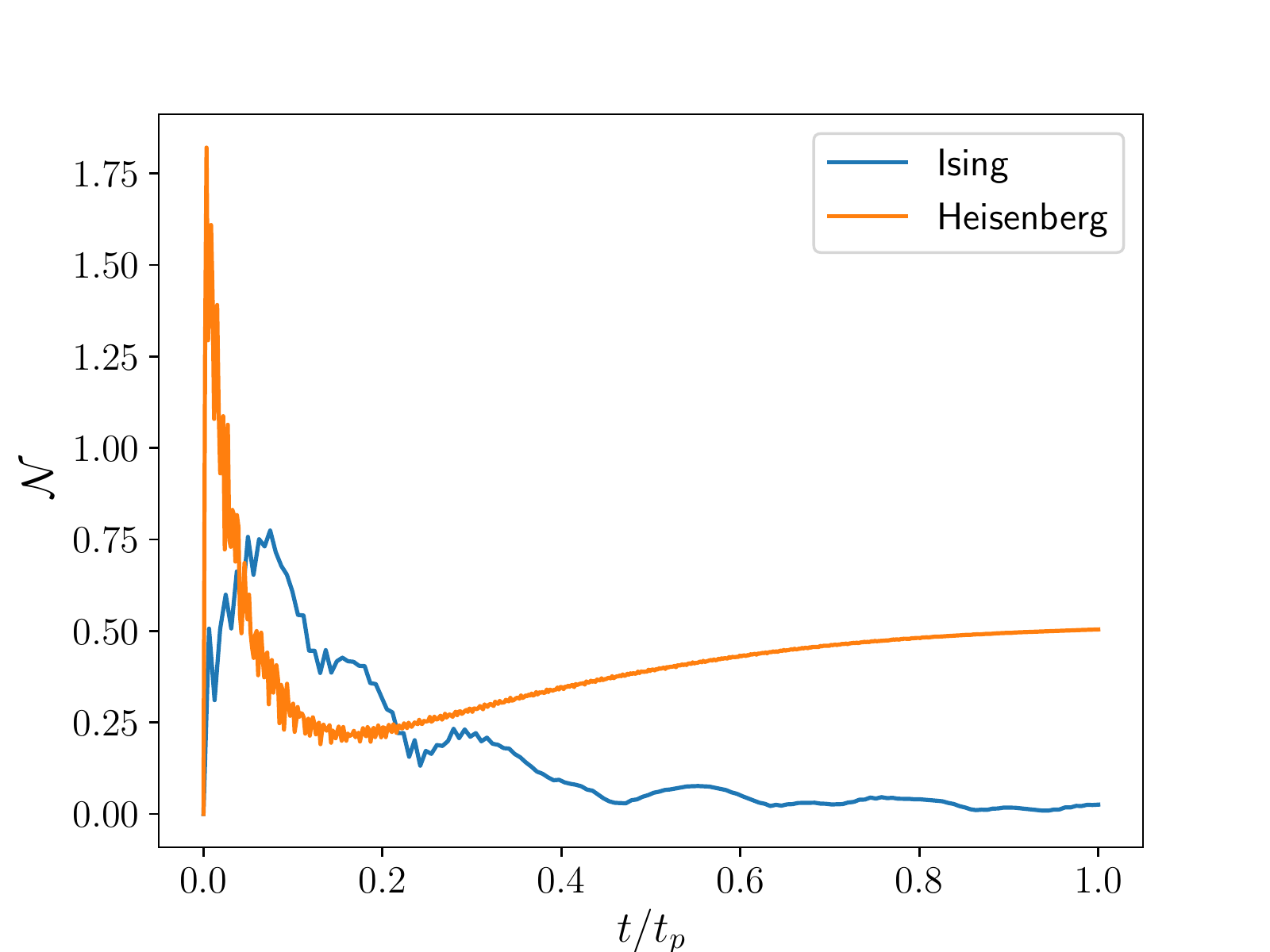}
\caption{Negativity as a measure of entanglement of the prepared states in time for a system of $N=6$ spins. The blue line corresponds to the Ising model ($J/g=5$) having a low negativity in the long time limit, whereas the orange line corresponds to the antiferromagnetic Heisenberg model with a highly entangled final state. The curves are shown for the optimized parameters for both cases leading to $\epsilon=0.2$.}
\label{fig:negativity}
\end{figure*}

As the ground state of the antiferromagnetic Heisenberg model is
highly entangled, this entanglement should be detectable within the
cooling dynamics. As an entanglement measure, we use the negativity
\begin{equation}
  \mathcal{N}(\rho) = \frac{\norm{\rho^{T_A}}_1-1}{2},
\end{equation}
where $\norm{.}_1$ refers to the trace norm and $\rho^{T_A}$ is the
partial transpose of $\rho$ with respect to the subsystem $A$
\cite{Vidal2002}. Here, we first trace out the bath spin and then take
half of the remaining system as the subsystem $A$. Figure
\ref{fig:negativity} shows the negativity of the Ising model and the
Heisenberg model as a function of time normalized with respect to the
total preparation time $t_p$. One can clearly see that the steady
state of the system exhibits large entanglement for the Heisenberg
model, \textcolor{black}{whereas} the Ising model is barely entangled. The initial spike
in the negativity can be attributed to the fact that typical
high-energy states follow a volume law for entanglement measures,
while ground states exhibit a weaker area law \cite{Eisert2010}. Note
that this initial entanglement is not useful for quantum information
processing tasks \cite{Gross2009}.

\newsavebox\myemptybib

\savebox\myemptybib{\parbox{\textwidth}{}}